\documentclass[
showpacs,floatfix,
twocolumn,%galley,
10pt,aip,apl,citeautoscript,superscriptaddress]{revtex4-1}

\usepackage{graphicx}
\usepackage{bm}
\usepackage{siunitx}

\begin{document}
\title{Synchronized single electron emission from dynamical quantum dots}

\author{P. Mirovsky}

\author{B. Kaestner}
\author{C. Leicht}
\author{A. C. Welker}
\author{T. Weimann}
\author{K. Pierz}
\author{H. W. Schumacher}

\affiliation{Physikalisch-Technische Bundesanstalt (PTB), Bundesallee 100, 38116 Braunschweig Germany.}

\date{November 02, 2010}

\begin{abstract}
We study synchronized quantized charge pumping through several dynamical quantum dots (QDs) driven by a single time modulated gate signal. We show that the main obstacle for synchronization being the lack of uniformity can be overcome by operating the QDs in the decay cascade regime. We discuss the mechanism responsible for lifting the stringent uniformity requirements. This enhanced functionality of dynamical QDs might find applications in nanoelectronics and quantum metrology.
\end{abstract}

%\pacs{73.63.Kv, 73.23.Hk, 73.21.La, 73.22.Dj}

\maketitle

Quantum dots (QDs) connected to leads have been a standard model system for many years to study single charges since the first proposal of single electronics~\cite{Averin1991}. Many of the promising applications in metrology and digital circuits make use of QDs with time varying confining potentials (dynamic QDs) to control \emph{single}-electron emission and capturing events.~\cite{likharev1PBI} To reach high integration densities and synchronized operation, it is important to minimize the number of individual device parameters. Due to intrinsic potential fluctuations, leading to a statistical distribution of the Coulomb blockade thresholds and energy barrier heights, even QDs with perfect geometry would require individual tuning settings. Typically, for dynamic QDs the number of independent variables grows, since the specific voltage modulation-amplitudes, offsets and phases add complexity to the system. Therefore even a simple parallel operation of dynamic QDs remains challenging.~\cite{keller2000, ebbecke5, wright2009} Recently, Maisi et al.~\cite{maisi2009} demonstrated the operation of ten single-electron turnstiles using only one global voltage modulation signal. The high reproducibility in these superconductor-metal-superconductor-based turnstile devices allowed single-electron operation in the \si{\mega\hertz} range with a low error margin using a common modulation signal. However present information technology devices are based on semiconductor materials which offer greater flexibility in tailoring the confinement potential landscape. This freedom to tune dynamic QDs by gate induced barrier modulation allows single electron pumping up to \si{\giga\hertz} frequencies.~\cite{blumenthal2007a} However, up to now the lack of reproducibility in the parameter range between different devices and cool-down cycles has rendered synchronization difficult.~\cite{wright2009} In this work we demonstrate synchronized emission of single electrons from multiple semiconductor QDs using one global driving gate. Synchronized emission is achieved by operating the QDs in the decay cascade~\cite{kaestner2010a} regime. Hence, we show that for this regime the requirements for uniformity between many devices can be significantly relaxed.

\begin{figure}
  \includegraphics[width=8.0cm]{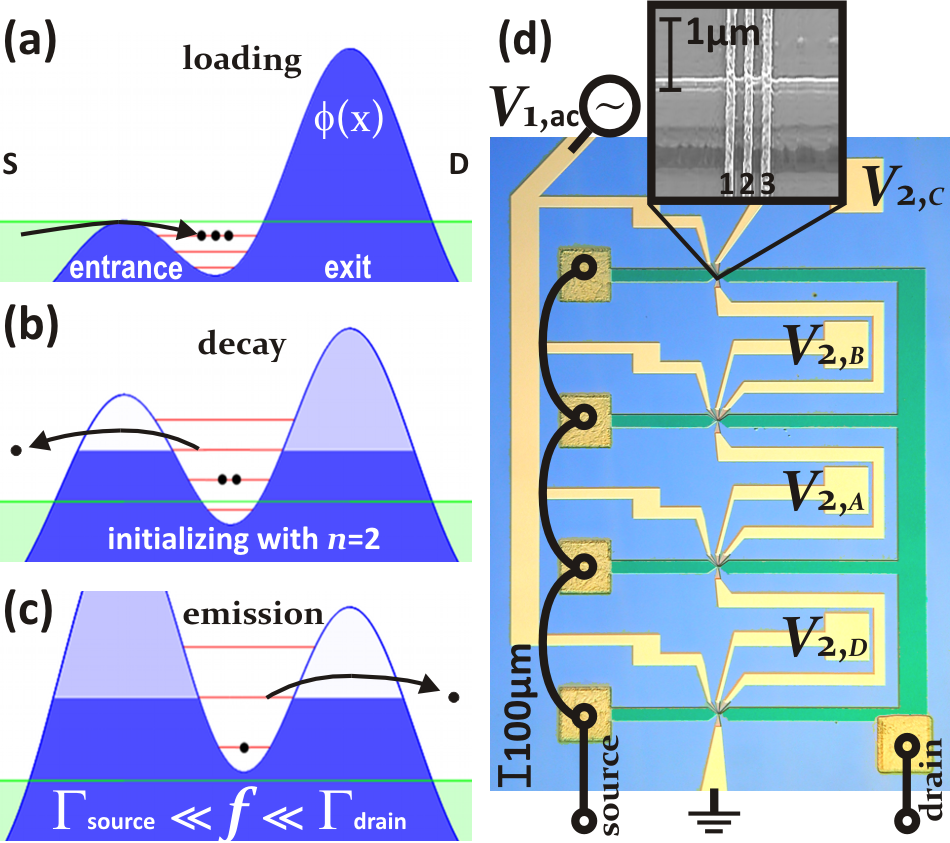}
  \caption{(Color online) Pumping scheme and device design. (a)-(c) show snap-shots of the pumping cycle phases. The blue curve represents a one-dimensional schematic of the confining potential landscape while the green line indicates the Fermi energy of the leads. In (b) and (c) the lighter-colored areas beneath the potential curve illustrate the different transparencies of the barriers for the escape events indicated by the arrows. Transparencies are linked to escape rates $\Gamma_{\text{source}}$, $\Gamma_{\text{drain}}$ to source (S) and drain (D), respectively. (d) optical microscope image with a scanning electron microscope inset. The \SI{100}{\nano\metre} wide gates (colored in yellow) are separated by \SI{150}{\nano\metre}. A voltage signal $V_{1,\text{\tiny ac}}$, composed of a sinusoidal signal with power $P_{\text{\tiny RF}}$ and the offset voltage $V_{1,\text{\tiny dc}}$, is applied to all entrance gates. The width of the etched constriction is about \SI{900}{\nano\metre} (channel with constriction colored in green).}
  \label{fig:devOp}
\end{figure}

The basic building block of our device is a dynamic QD, where the periodically varying confining potential is defined by two voltage parameters. A schematic one-dimensional representation of the potential energy landscape $\phi(x)$ is shown in Fig.~\ref{fig:devOp}. The different stages of the cycle leading to controlled single electron emission based on the decay cascade are illustrated in Figs.~\ref{fig:devOp}(a)-(c). During the first part of the cycle the entrance barrier is lowered allowing electrons from source to enter the QD (Fig.~\ref{fig:devOp}(a)). Afterwards, while the entrance barrier increases, the QD is raised energetically above the Fermi level of the source allowing non-equilibrium relaxation from the QD (decay cascade, Fig.~\ref{fig:devOp}(b)). The barrier finally becomes sufficiently opaque to isolate a number of electrons, $n$, from the source, where back tunneling becomes very unlikely within the cycle period. Here time scale separation in the escape rates between subsequent charge states results in a low-dispersion initialization of the QD~\cite{kaestner2010a}. During the second stage of the pumping cycle the entrance barrier as well as the energy of the QD continues to raise such that the exit barrier becomes transparent and electrons start to tunnel to drain (Fig.~\ref{fig:devOp}(c)). If at this point the modulation amplitude of the entrance barrier voltage is too small, the exit barrier is not sufficiently transparent for all electrons to be emitted during the cycle period. Therefore, there is a requirement for a minimal modulation amplitude that assures a complete draining of the QD. Once all electrons are emitted a so called \emph{quantized} current of $I = n\,e\,f$ is generated, where $e$ is the elementary charge and $f$ the frequency at which the cycle is repeated. In principle, increasing the modulation amplitude further does not have any effect on $I$. This is one of the most important features of the decay cascade mechanism and responsible for lifting the stringent uniformity requirements for synchronized single electron emission from the dynamical QD.

The pumping scheme has been implemented here in an n-type AlGaAs/GaAs heterostructure and is shown in Fig.~\ref{fig:devOp}(d). An arrangement of four identical structures can be seen allowing synchronous operation as described below. The inset shows the individual device consisting of a constriction of about \SI{900}{nm} width that was defined by wet-etching the dopant layer of the heterostructure. The constriction is crossed by three Ti-Au top gates. A QD between the first two gates is formed by applying sufficiently large negative dc voltages $V_{1,\text{\tiny dc}}$ and $V_{2,i}$ to the entrance and exit gate, respectively (the index $i$ labels the individual pump devices $A$, $B$, $C$, $D$). The third gate was not used in this experiment and therefore set to ground. An additional sinusoidal signal of power $P_{\text{\tiny RF}}$, adding up with $V_{1,\text{\tiny dc}}$ to $V_{1,\text{\tiny ac}}$, is coupled to the entrance gate in order to generate the required time dependent confinement potential.

As it can be seen from Fig.~\ref{fig:devOp}(d) the entrance gates of all four pump devices are lithographically connected together while the exit gate can be tuned individually. In addition, all source contacts have been bonded together. In order to characterize a single pump device the exit gate voltage of all other pumps needs to be set to a sufficiently negative voltage.

\begin{figure}
  \includegraphics[width=8.0cm,height=7.0cm]{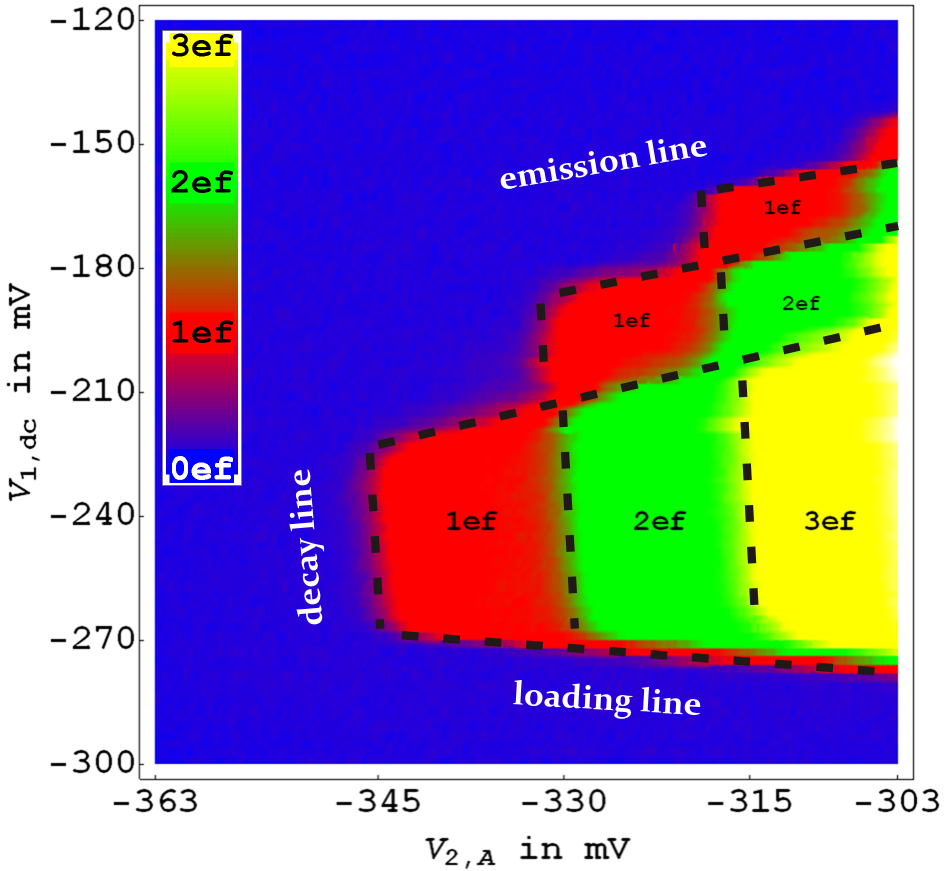}
  \caption{(Color online) Pumped current as a function of the entrance- and the exit-gate for pump $A$. The measurement was performed at $P = -14\,\text{dBm}$ at $f = \SI{550}{\mega\hertz}$. The resulting current $I$ is color-coded resulting in quantized pumping plateaus. The edges are emphasized with dashed black lines parallel to the corresponding description (see main text).}
  \label{fig:quantized}
\end{figure}

Fig.~\ref{fig:quantized} shows the characterization of the pump with the label $i = A$. It illustrates that by varying $V_{1,\text{\tiny dc}}$ and $V_{2,A}$ pump $A$ can be switched between regions of quantized current generation. Within the blue region pumping is not possible. Red areas indicate a current of $I = 1\,e\,f$, while green corresponds to $2\,e\,f$ and yellow to $3\,e\,f$. The borders of the pumping region have been discussed in detail e.g. in Ref.~\onlinecite{Leicht2009}. In summary, the lower step edge labeled \emph{loading line} is defined by the transition when electrons start to enter the QD during the loading phase (see Fig.~\ref{fig:devOp}). The left step edge (\emph{decay line}) of each plateau emerges due to different final populations of the QD at the end of the initialization phase. The QD state reaches its maximal energy during the emission phase. An insufficiently negative entrance gate voltage $V_{1,\text{\tiny dc}}$ leads to an incomplete draining of the QD, resulting in the upper step edge (\emph{emission line}). Hence, the current is determined predominantly by the decay cascade process as long as $V_{1,\text{\tiny dc}}$ is set with sufficient distances to the loading and emission line. It follows from the arguments above that increasing the power $P_{\text{\tiny RF}}$ enhances the robustness in $V_{1,\text{\tiny dc}}$, i.e. it extends the decay cascade dominated transport by shifting the loading and emission line apart, while in principle the distances between the decay lines would not change at all. This behavior has been demonstrated experimentally in Ref.~\onlinecite{kaestner2008} and allows us to expand the decay cascade dominated transport until all pump devices can be operated in this regime at the same working point, $V_{1,\text{\tiny dc}}$. The fixed robustness in $V_{2,i}$, mainly defined by the back tunneling cascade process, results in $V_{2,i}$ as the only remaining parameter that has to be tuned individually for each pump device.

\begin{figure}
  \includegraphics[width=8.0cm]{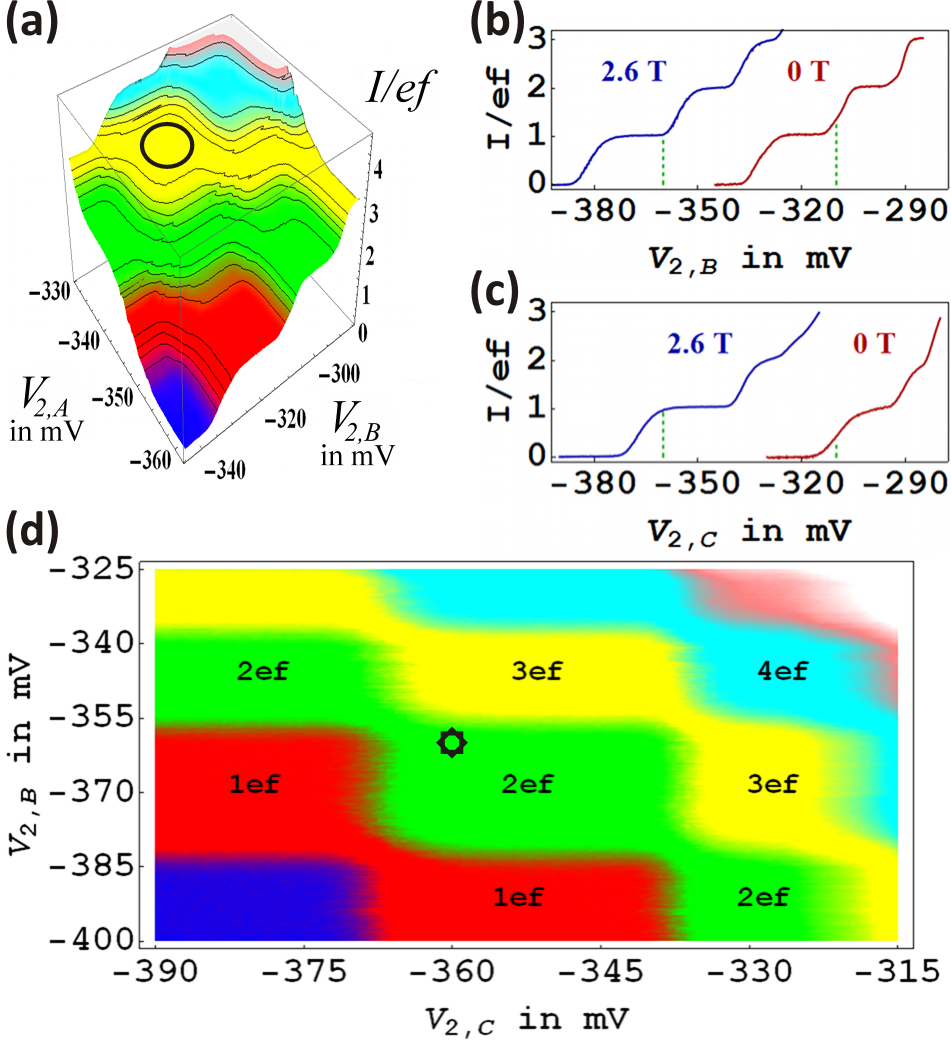}
  \caption{(Color online) Pumped current as a function of the exit-gate voltages. In (a) all three working pumps are operated at $P_{\text{\tiny RF}} = -14\,\text{dBm}$. The exit voltage of pump $C$ is linked via $V_{2,C} = V_{2,A} + \SI{40}{\milli\volt}$ to the pump $A$. (b) and (c) display line scans for $\text{B}=\SI{2.6}{\tesla}$ and without field along $V_{2,B}$ and $V_{2,C}$, respectively. The vertical lines at \SI{-360}{\milli\volt} and \SI{-310}{\milli\volt} clarify that the plateau enhancement leads to an overlap. Under magnetic field operation $P_{\text{\tiny RF}} = -10\,\text{dBm}$ was required to remain in the decay cascade regime. (d) shows synchronized operation of pumps $B$ and $C$ under the same conditions as in (b) and (c) with $\text{B} = \SI{2.6}{\tesla}$. Pump $A$ was blocked with $V_{2,A} = \SI{-450}{\milli\volt}$. The position of the black marker corresponds to $V_{2,B} = V_{2,C} = \SI{-360}{\milli\volt}$. All above measurements were performed at  $f = \SI{550}{\mega\hertz}$ and $V_{1,\text{\tiny dc}} = \SI{-240}{\milli\volt}$.}
  \label{fig:Plateaus}
\end{figure}

Fig.~\ref{fig:Plateaus} assembles the results of synchronized operation of up to three pumps. The measurements were performed in a $^{3}\text{He}$ cryostat at a temperature below \SI{500}{\milli\kelvin}. Although all pumps have nominally the same geometry and are made during the same fabrication process, we found that pump $D$ is not working, possibly due to lithographically caused gate failure, while the others show a strong variation of the plateau regions. While the plateau of pump $A$ is centered around $V_{1,\text{\tiny dc}} = \SI{-245}{\milli\volt}$ (see Fig.~\ref{fig:quantized}) a similar measurement resulted in $V_{1,\text{\tiny dc}} = \SI{-240}{\milli\volt}$ for pump $B$ and $V_{1,\text{\tiny dc}} = \SI{-200}{\milli\volt}$ for pump $C$ (not shown). The resulting overlap at $P_{\text{\tiny RF}} = -14\,\text{dBm}$ and $f = \SI{550}{\mega\hertz}$ is approximately \SI{20}{\milli\volt} centered at \SI{-240}{\milli\volt} and comparable to the $V_{2,i}$-plateau width. For smaller overlaps of the decay cascade regions we expect insufficient draining or loading of the QD to be the dominant accuracy limiting factor. To demonstrate synchronized pumping operation we set the global offset voltage of all entrance gates to $V_{1,\text{\tiny dc}} = \SI{-240}{\milli\volt}$ and vary $V_{2,B}$ and $V_{2,C}$, and link $V_{2,A}$ to $V_{2,C}$ via $V_{2,C} = V_{2,A} + \SI{40}{\milli\volt}$ to have a good overlap of both pumping plateaus. In Fig.~\ref{fig:Plateaus}(a) the yellow region marked with a circle and the corresponding current of $3\,e\,f$ equals a current of \SI{264.36}{\pico\ampere} produced by all three pumps, each pumping $1\,e\,f$.

As discussed above, the fixed robustness in $V_{2,i}$ means that there remains one dc parameter that has to be tuned individually for each pump device. However, it has been found empirically that a perpendicular magnetic field may enhance the robustness in $V_{2,i}$ (the origin is not yet completely understood).~\cite{kaestner2009a, wright2008} Therefore, a perpendicular magnetic field of $\text{B} = \SI{2.6}{\tesla}$ has been applied, modifying the individual pumping characteristics as shown in Fig.~\ref{fig:Plateaus}(b) and (c). It results not only in shifting of the plateaus but also enhancement of the robustness. Note, that other methods for plateau enhancement can be envisaged, such as increasing the charging energy.~\cite{kaestner2010a} In this case, however, magnetic field application enhances the robustness of pumps $B$ and $C$ in $V_{2,i}$ sufficiently so that synchronized emission can be achieved applying the \emph{same} exit gate voltage of $V_{2,B} = V_{2,C} = \SI{-360}{\milli\volt}$. The corresponding synchronized operation of pumps $B$ and $C$ with pump $A$ set to zero pumped current is demonstrated in Fig.~\ref{fig:Plateaus}(d). Varying the exit gate voltages $V_{2,B}$ and $V_{2,C}$, one obtains a checker board pattern. The green area in the center of the plot arises from the combination of pump $B$ and $C$ each pumping one electron per cycle, adding up to a current of $2\,e\,f$. The possibility to use two common control voltages only is indicated by the black marker.

In conclusion, we have demonstrated that operating dynamic QDs in the decay cascade regime allows synchronized emission of a controllable number of electrons, triggered by one global driving gate. Furthermore, two QDs have been tuned simultaneously into the single-electron emission regime by only using two control gates in total. Thus this type of dynamical QD with its enhanced functionality might find applications in nanoelectronics, and in particular in quantum metrology as building block for macroscopic current sources with quantum limited precision.

We thank V. Kashcheyevs and F. Hohls for useful discussions and H. Marx and U. Becker for their technical assistance. This work has been supported by DFG and EURAMET joint research project with European Community's 7th Framework Programme, ERANET Plus under Grant Agreement No. 217257. C.L. has been supported by International Graduate School of Metrology, Braunschweig. 
% \comm{Funding by EuroMagNET under the EU Contract No. RII3-CT-2004- 506239 is acknowledged. DFG}
\vspace{-3ex}
\small
%\bibliography{literatureJabRef}

\end{document}